\documentclass[useAMS,usenatbib,usegraphicx]{mn2e}
\usepackage{epsfig}
\usepackage{amsmath}
\usepackage{natbib,graphicx,color}
\usepackage[normalem]{ulem} 
\def\apj {ApJ}

\def\apjs {ApJS}
\def\aj {AJ}

\def\aap {A\& A}
\def\mnras {MNRAS}

\title{Structure and dynamics of the supercluster of galaxies SC0028-0005}

\author[O'Mill et. al.]
{Ana Laura O'Mill$^{1,5}$,
Dominique Proust$^{2}$,
Hugo V. Capelato$^{3,4}$,
Mirian Castejon$^{1}$,
\newauthor Eduardo S. Cypriano$^{1}$,
Gast\~ao B. Lima Neto $^{1}$ and
Laerte Sodr\'e Jr.$^{1}$\\
$^1$ Departamento de Astronomia, Instituto de Astronomia, Geof\'\i sica
e Ci\^encias Atmosf\'ericas da USP,\\ ~~~Rua do Mat\~ao 1226, Cidade
Universit\'aria, 05508-090, S\~ao Paulo, Brazil.\\
$^2$ Observatoire de Paris-Meudon, GEPI, F92195 MEUDON, France.\\
$^3$ Divis\~ao de Astrof\'\i sica, INPE/MCT, 12227-010, S\~ao Jos\'e dos Campos,
S\~ao Paulo, Brazil.\\
$^4$ N\'ucleo de Astrof\'\i sica Te\'orica, Universidade Cruzeiro do Sul, Rua Galv\~ao Bueno,868,CEP 01506-000, S\~ao Paulo, Brazil.\\
$^5$ Instituto de Astronom\'\i a Te\'orica y Experimental, CONICET-UNC, Laprida 922, C\'ordoba, Argentina.\\ 
}

\date{\today}

\begin{document}
\pagerange{\pageref{firstpage}--\pageref{lastpage}}

\maketitle

\label{firstpage}

\begin{abstract}

% Context
According to the standard cosmological scenario, superclusters are objects that have just passed the turn around point and are collapsing. The dynamics of very few superclusters have been analysed up to now.
%
% Aims
In this paper we study the supercluster SC0028-0005, at redshift 0.22, identify the most prominent groups and/or clusters that make up the supercluster, and investigate the dynamic state of this structure.
%
% Method
For the membership identification, we have used photometric and spectroscopic data from SDSS-DR10, finding 6 main structures in a flat spatial distribution. We have also used a deep multi-band observation with MegaCam/CFHT to estimate de mass distribution through the weak-lensing effect. For the dynamical analysis, we have determined the relative distances along the line of sight within the supercluster using the Fundamental Plane of early-type galaxies.
Finally, we have computed the peculiar velocities of each of the main structures.
%
% Results
The 3D distribution suggests that SC0028-005 is indeed a collapsing supercluster, supporting the formation scenario of these structures. Using the spherical collapse model, we estimate that the mass within $r = 10$~Mpc should lie between 4 and $16 \times 10^{15} M_\odot$. The farthest detected members of the supercluster suggest that within $\sim 60$~Mpc the density contrast is $\delta \sim 3$ with respect to the critical density at $z=0.22$, implying a total mass of $\sim 4.6$--$16 \times 10^{17} M_\odot$, most of which in the form of low-mass galaxy groups or smaller substructures.

\end{abstract}


%%%%%%%%%%%%%%%%%%%%%%%%%%%%%%%%%%%%%%%%%%%%%%%%%%%%%%%%%%%%%%%%%%%%%%%%%%%%%%%%%%%%%%%%%%%%%%

\section{Introduction}

In the hierarchical paradigm of structure formation, the smallest structures are the 
first to collapse and virialize. They later get collected into progressively larger
structures, each of which goes through the same stages of collapse and virialization.
At present, the largest collapsed structures are clusters of galaxies, whereas
superclusters are expected to be in the stage of gravitational collapse, at least in their inner tens of Mpc 
\citep{Reisenegger00,Batiste13, Merluzzi15}. In the currently favoured cosmological
scenario, dominated by a cosmological constant or another form of ``dark energy'',
superclusters are starting to recede from each other at an accelerated rate, which will
not allow them to get collected into even larger structures. Therefore, they are the
largest structures that will ever collapse and virialize \citep[e.g.][]{Nagamine03, Busha03, 
Dunner06, Dunner07, Proust06, Rines13}. 
If this scenario 
is correct, present superclusters play a pivotal role in our understanding of the 
evolution of the universe. Thus, it is very important to determine their quantitative 
properties such as masses, sizes, and densities reliably, as well as testing this scenario 
as best as possible. Since superclusters have not yet collapsed and virialized, they do 
not stand out against the background density field as clearly as individual galaxies or
even clusters of galaxies (in terms of mass, optical light or X-ray emission), and
equilibrium techniques used to determine the masses of clusters (hydrostatic equilibrium
of the gas and virial equilibrium of the galaxies) do not apply to them.

The study of superclusters has a long history, beginning with works such as the identification of our own Virgo supercluster \citep{deVaucouleurs53} and the statistical description of ``clustering of second order'' \citep[which are known nowadays as the superclusters;][]{Neyman56}. Other studies addressed the identification, distribution and characterization of these objects \citep[e.g.][]{Joeveer78, Tago84, Einasto84, Einasto94, Zucca93}, establishing that they are separated by large voids and clusters inside them are usually organized in chain-like structures.

Some studies considered the morphology of superclusters, based on the spatial distribution of their galaxies. 
\citet{Einasto07, Einasto11} analyzed a sample drawn from SDSS  DR7, finding two main morphological types: filaments and others with a more complex, multibranch fine structure. The wide morphological variety of superclusters let \citet{Einasto11} to suggest that their evolution have been dissimilar.
\citet{CostaDuarte11} used a kernel-based density 
field method to identify the superclusters and Minkowski Functionals to quantify their shape. 
They found that filaments and pancakes represent distinct morphological classes of 
superclusters in the Universe. The filaments tend to be richer, more luminous and larger 
than pancakes. It is then plausible to think that pancakes evolve towards filaments. 

While there is plenty of morphological analysis, dynamical studies of galaxy superclusters are more scarce.
In the hierarchical model of structure formation we expect that galaxy clusters
form through accretion of galaxies and groups or merger with other clusters in
the environment of superclusters. 
In the case of a supercluster with several clusters, it is possible that massive clusters grow through the gravitational collapse of the central parts of superclusters. Therefore, superclusters are still far from equilibrium today. The collapse scenario predicts that, among the 
clusters in a supercluster, those with higher observed redshifts are falling
in to the centre from the front side (and thus are closer to us), while
those with lower observed redshifts are falling in from the back 
(and are therefore more distant). Consequently, in the gravitationally
collapsing region, the Hubble relation is reversed. This effect is indeed present 
in the core of the Shapley Supercluster \citep{Proust06, Ragone06, Dunner07}, at 
$z \simeq 0.04$. \citet{Batiste13} made a dynamical analysis of the Corona Borealis 
supercluster at $z \simeq 0.07$ using data from the SDSS. They find this supercluster 
has broken from the Hubble flow and, based on dynamical simulations, conclude that a significant 
fraction of mass should reside outside the clusters comprising the supercluster. Recently, \citet{Tully14}
have used peculiar velocities to identify and describe a large and massive structure, Laniakea, comprising most of the galaxies in the local universe.

%\textbf{The picture of gravitationally collapsing central parts of superclusters is required by 
%the present cosmological model, but it has been tested only at low directly by observations at redshifts. Indeed, previous dynamical studies, as those described above and in Sec.~\ref{sec:dynamical}, dealt with superclusters below $z=0.1$.}
The picture of gravitationally collapsing central parts of superclusters is required by 
the present cosmological model, but it has been tested only at low redshifts. Indeed, previous dynamical studies, as those described above and in Sec.~\ref{sec:dynamical}, dealt with superclusters below $z=0.1$.
In this work, we analyze the supercluster
SC0028-0005, at $z=0.22$, in order to test the collapse scenario for structures at intermediate redshift.
This paper is organized as follows. In Section~\ref{sec:description} we describe what is known about this supercluster and what is the data that we will analyze in this paper. In Section~\ref{sec:identify} we present an identification of the supercluster members. Section \ref{sec:lens} contains a weak-lensing analysis of the substructures found in the supercluster. In Section~\ref{sec:dynamical}  we show that
the dynamical behaviour of the supercluster components is consistent with a collapsing scenario. For this analysis 
we use the Fundamental Plane of early-type galaxies to obtain relative distances and peculiar velocities of these components. Finally, our results are summarized in Sec.~\ref{sec:conclusion}.
Throughout this paper we adopt, when necessary, a $\Lambda$CDM cosmological model with $\Omega_M = 0.30$, $\Omega_{\Lambda} = 0.70$ and $H_0=70 ~ h_{70}\,$km~s$^{-1}$~Mpc$^{-1}$.

%%%%%%%%%%%%%%%%%%%%%%%%%%%%%%%%%%%%%%%%%%%%%%%%%%%%%%%%%%%%%%%%%%%%%%%%%%%%%%%%%%%%%%%%%%%%%%

\section{The supercluster SC0028-0005}\label{sec:description}

In this study we focus on the galaxy supercluster SC0028-0005 (hereafter SC0028 for simplicity) at $\alpha=$ 00h28m and $\delta= -00^\circ 05^\prime$ (J2000.0). It is in the catalogue of \citet{Basilakos03}, obtained from the Sloan Digital Sky Survey (SDSS) ``Cut \& Enhance'' cluster catalogue \citep{Goto02} by applying a percolation radius $R_{\rm pc}= 26 h^{-1}$~Mpc. Among the 57 superclusters of this catalogue, SC0028 is the number 38, with an assigned redshift $z = 0.197$, and is classified as a \textit{filament} by using the shape finder estimator of \citet{Basilakos03}.  This structure was chosen for this study because it was not too complex in the \citet{Goto02} catalogue, containing only 3 galaxy clusters with relative distances of 4 to $10 h^{-1}$~Mpc between them, what was suggestive of the presence of strong peculiar motions.

%--------------------------------------------------------------------------

\begin{figure*}
\includegraphics[width=14cm]{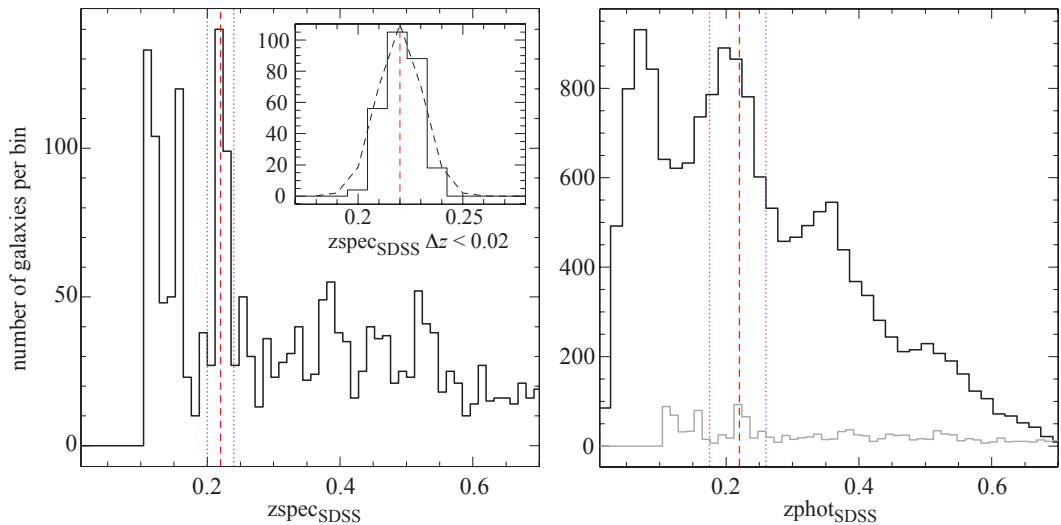}
\caption[]{
Redshift distribution within an area of 1.2 degrees radius centred on SC0028. \textsc{Left panel}: Spectroscopic redshift distribution. The inset shows the redshift distribution of galaxies in 
the $\Delta z < 0.02$ interval around the supercluster mean redshift, as well as a Gaussian fit of this distribution.
\textsc{Right panel}: Photometric redshifts. The gray histogram correspond to the redshift sample, reported here for easy comparison. In both panels,
the dashed vertical lines correspond to the supercluster mean redshift and the magenta dotted lines show 
the galaxies within a redshift interval around the mean of $\Delta z < 0.02$ for the S1 sample and $\Delta z_{phot} < 0.04$ 
for the S2 sample. 
}
\label{f1}
\end{figure*}

%--------------------------------------------------------------------------

\begin{figure*}
\includegraphics[width=15cm]{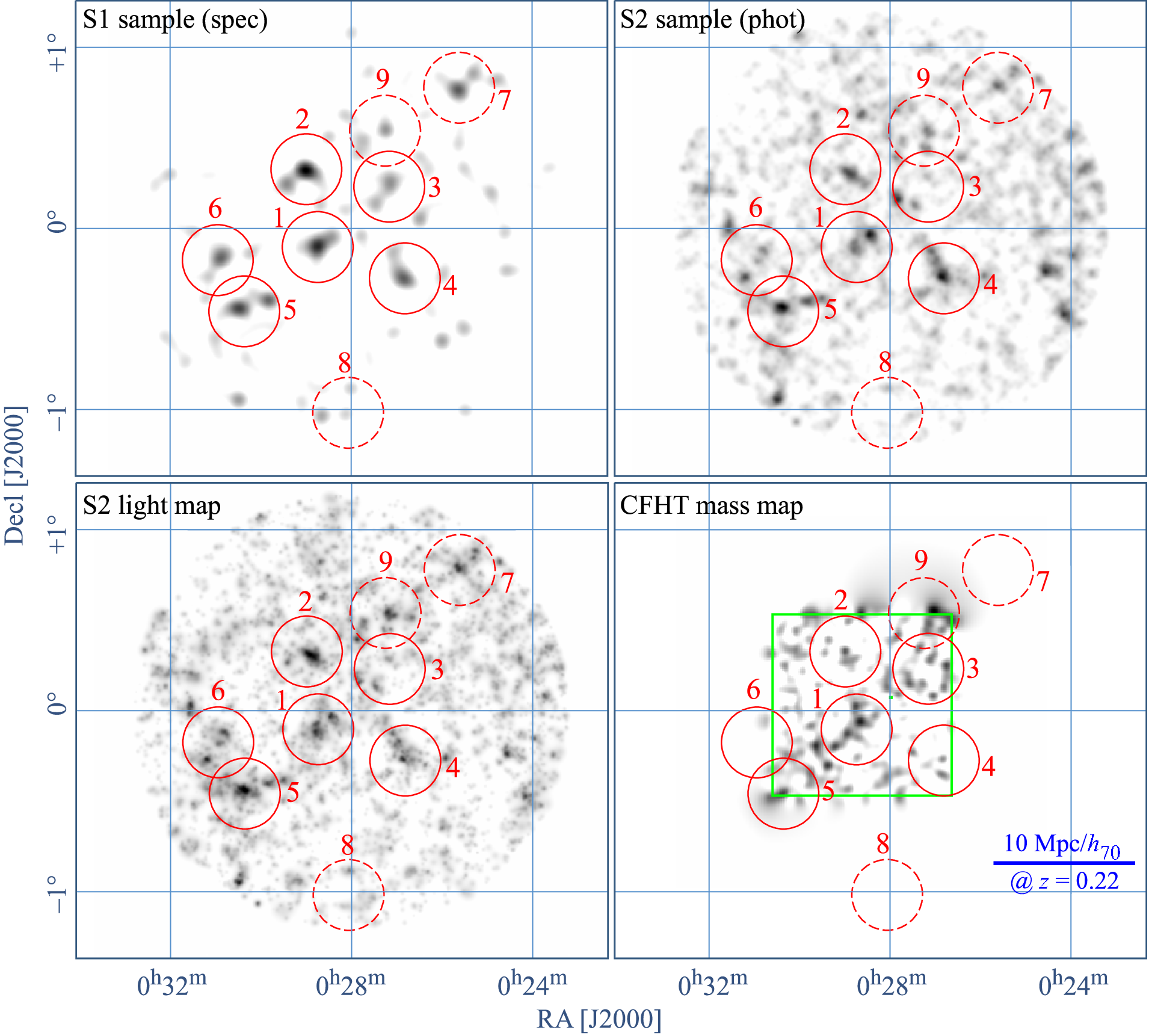}
\caption[]{Substructures in the supercluster. 
\textsc{Top-left}: Surface density map of supercluster members in the S1 (spectroscopic) sample.
\textsc{Top-right}: Surface density map of supercluster members in the S2 (photometric) sample.
\textsc{Botton-left}: Projected light distribution of the galaxies within photometric redshift $\Delta z < 0.04$ of the supercluster mean spectroscopic redshift, $z=0.22$. 
\textsc{Bottom-right}: Weak-lensing recovered mass distribution (See Section~\ref{sec:lens} for details). The green square corresponds to the 1 square degree MegaCam/CFHT field of view.
In all panels, the red circles ($R = 11.7$ arcmin) correspond to maximum projected density regions (based on the S1 sample), recentered at the location of the corresponding brightest galaxy in each maximum density box. 
The circles with solid lines are the substructures that will be considered in this study (see text for details).
}
\label{f2}
\end{figure*}

\subsection{Spectroscopic and photometric samples}
\label{sec:z_samples}

We selected from the Data Release 10 of Sloan Digital 
Sky Survey\footnote{https://www.sdss3.org/dr10/} (SDSS-DR10) a sample of galaxies with spectroscopic and 
photometric data within an area of radius 1.2 degrees, centred on the supercluster.

SDSS DR10 covers an additional 3100 sq. degree of sky over the previous release and
includes spectra obtained with the new spectrographs APOGEE and BOSS, whose sky coverage includes the region of SC0028. 

The spectroscopic and photometric data sets  were extracted from the \texttt{SpecObjAll} and 
\texttt{PhotoObj} tables of the \texttt{CasJobs}\footnote{http://skyserver.sdss3.org/CasJobs/} database.
From the \texttt{PhotoObj} table we downloaded the \texttt{objID}, the coordinates, the model $g$ and $r$ magnitudes, the Galaxy extinction values, the effective radii derived from the de Vaucouleurs 
profile fit, as well as  model magnitudes and 
axial ratios of the de Vaucouleurs 
fits, and the photometric redshifts. 
From the \texttt{SpecObjAll} table we downloaded the spectroscopic redshifts and the galaxy central velocity dispersions plus the respective errors.

From this data we selected two galaxy samples.
The first sample %, hereafter S1,
comprises 4921 galaxies with spectroscopic 
redshift. This sample is employed to identify the supercluster members and substructures, as well as in the dynamical analysis of Section~\ref{sec:dynamical}. 
  
The second sample %, S2,
contains 35757 galaxies with photometric redshifts (including the galaxies of the previous spectral sample),
selected from the \texttt{PhotoObj} table. This sample allowed an
analysis of the photometric properties of the substructures 
identified spectroscopically and is also useful for their detection.

\subsection{CFHT Imaging}
\label{sec:CFHT}

The supercluster field was observed with CFHT/MegaCam in imaging mode, mostly for the weak-lensing analysis described in Section \ref{sec:lens}. We got deep, good quality {\it g}, {\it r}, {\it i} images, whose main features are shown in Table \ref{cfht:phot}. 

The whole data processing (bias and overscan subtraction, flatfielding and sky subtraction), image combination and catalogue extraction with SExtractor \citep{SExtractor} has been done by the Terapix\footnote{http://terapix.iap.fr/} team, using the procedures described in \citet{Terapix}.

The total integrations were obtained with a series of $\sim$ 10 min. exposures per filter, with small position offsets, for the coverage of the cap between the camera CCDs. As a result, the final combined image in each band covers an area slightly smaller than one square degree (MegaCam's full field of view).

\begin{table}
\caption{CFHT Imaging characteristics}
\label{cfht:phot}
\begin{tabular}{|cccc|}
\hline
\hline
Band & Exposure & Seeing  & Completness\\
     & (h)      & (arcsec)& (AB mag.)\\
\hline
g & 1.5 & 0.52 & 24.5 \\
r & 3.3 & 0.45 & 24.6 \\
i & 2.5 & 0.45 & 24.3 \\
\hline
\hline
\end{tabular}
\end{table}

%%%%%%%%%%%%%%%%%%%%%%%%%%%%%%%%%%%%%%%%%%%%%%%%%%%%%%%%%%%%%%%%%%%%%%%%%%%%%%%%%%%%

\section{Identifying the Supercluster members}\label{sec:identify}

%%%%%%%%%%%%%%%%%%%%%%%%%%%%%%%%%%%%%%%%%%%%%%%%%%%%%%%%%%%%%%%%%%%%%%%%%%%%%%%%%%%%

\subsection{Selecting galaxies}

With the aim of selecting galaxies which are members of the supercluster, we start
by defining $\Delta z$ as the absolute difference between a given redshift 
and the supercluster mean redshift. For galaxies with spectroscopic redshifts 
we consider as members those with $\Delta z < 0.02$. We call this sample S1. 
For those with photometric redshifts only, due to the larger uncertainties, 
we adopted $\Delta z_{phot} < 0.04$ (about twice the SDSS photometric redshift
standard deviation). This we call the S2 sample.

Left and right panels of Figure \ref{f1} show the resulting redshift distribution of the galaxies in the spectroscopic and photometric samples (Section~\ref{sec:z_samples}), respectively. The peak around $z \simeq 0.22$ in these figures corresponds to the supercluster.

We selected galaxies from the spectroscopic sample which are supercluster members using the following approach. We preselected all galaxies in the range $0.195<z<0.245$, a redshift interval large enough to not exclude possible members, and small enough to avoid most fore/background galaxies. We fitted a Gaussian to the resulting redshift distribution, obtaining  an average redshift of $\overline{z}=0.220 \pm 0.001$. Around this value we selected the spectroscopic sample, S1, using $\Delta z < 0.02$, resulting in a total of 271 galaxies. For the photometric sample we adopted $\Delta z_{phot} < 0.04$, obtaining 3549 galaxies. In both panels of Figure \ref{f1} the magenta dotted lines bracket the membership ranges.

%--------------------------------------------------------------------------

\begin{table}
\caption[]{Structures within the supercluster: 
Identification, coordinates, mean redshift and red fraction.\\}
  \begin{tabular}{lcccc}
  \hline
  \hline
Id &  $\alpha$ (J2000.0) & $\delta$ (J2000.0) & $z$ & fraction of red  \\
            &  [h m s]   &   [$^{\circ}$ \ ' \ "] &   & galaxies [$\%$]   \\
\hline
1      & 00 28 44.28  &  $-00 05 47.16$   & 0.218$\pm$ 0.012& 78 \\ 
2      & 00 28 59.35  &  $+00 20 00.64$   & 0.222$\pm$0.015 & 66 \\
3      & 00 27 09.43  &  $+00 14 13.95$   & 0.217$\pm$0.017& 55 \\
4      & 00 26 48.78  &  $-00 16 09.73$   & 0.226$\pm $0.017& 69 \\
5$^*$  & 00 30 21.65  &  $-00 27 05.62$   & 0.225$\pm$ 0.015& 61 \\
6$^*$  & 00 30 56.81  &  $-00 10 07.15$   & 0.225$\pm$ 0.017& 57 \\
7      & 00 25 36.75  &  $+00 47 01.09$   & 0.220$\pm$0.017& 38 \\
8      & 00 28 03.93  &  $-01 00 42.28$   & 0.217$\pm $0.015& 30 \\
9     & 00 27 14.99   &  $+00 32 46.67$   & 0.217$\pm$ 0.015 & 46 \\
 \hline
SC & 00 28 00.00  &  -00 05 00.00   & 0.220$\pm0.001$ & 81$^{**}$ \\
\hline
\hline
\label{t1}
\end{tabular}
{\bfseries Note$^*$:} There is some overlap between regions 5 and 6 (Figure \ref{f2}). After checking the (photometric) redshift of the 6 galaxies in the overlap region we decided that this area should belong to the structure in region 5. \\
{\bfseries Note$^{**}$:} The fraction of red galaxies was determined with the S2 sample. 
\end{table}

%--------------------------------------------------------------------------
\subsection{Identifying substructures}

In order to identify substructures in our spectroscopic sample, we present a density map 
of the S1 sample in the upper left panel of Figure \ref{f2}. 
This map was made by counting the number of galaxies in equal size square cells 
with 15 arcsec on a side. 
We found 9 well defined peaks in this density distribution which are most likely 
galaxy groups or clusters candidate members of SC0028. 
The brightest galaxy in each peak was taken as the group/cluster centre, and all galaxies 
in the sample within a 2.5 $h_{70}^{-1}$ Mpc radius ($\sim 11.7$~arcmin) of this centre were considered as group/cluster members. 
In this panel, the circles show the region around each peak ascribed to a certain 
group or cluster. Table \ref{t1} presents the identification, coordinates, and mean 
redshift of each of these substructures. Notice that they may be groups or galaxy clusters, but hereafter we will
call them substructures.

The upper right panel of Figure \ref{f2} is similar to the left panel, but uses the larger S2 sample (the photometric one) instead of the S1 sample.

The light map (Figure \ref{f2}, lower left panel) was made by strongly smoothing the S2 sample of galaxies by a bidimensional Gaussian with intensity proportional to the galaxy luminosity and width proportional to their half-luminosity radius, $\sigma = 100 R_{50}$. In this way, we can take into account the different apparent sizes of the galaxies.
This luminosity field was then binned in a 2D grid with $15 \times 15$~arcsec$^2$ cells, producing an image. Since the mean $R_{50}$ is $\sim 1.1$~arcsec, the typical galaxy luminosity was spread in a radius of $\sim 7$~pixel ($1\sigma$), which we found adequate to describe the substructures in the light distribution.
This map is in qualitative agreement with the S1 density map, as there are peaks in most regions previously identified (solid line circles in figure~\ref{f2}).

The bottom right panel of figure~\ref{f2} shows the weak-lensing recovered projected mass distribution (see Section~\ref{sec:lens}). The same structures identified in the previous maps are also present in this mass map, within the CFHT field of view (green square in the figure). This reassures that these substructures are not projection effects, but rather actual mass concentrations. Substructure 3 is the only exception, since it is detected in redshift space, upper left panel, but not clearly visible in any of the others. This probably means that substructure 3 is the least massive region of them all.

In figure~\ref{f2} we can see that areas 5 and 6 overlap.  
After analyzing the photometric redshifts of the 6 galaxies in the area in common, we decided to consider them as a part of group/cluster 5.

Some of those substructures might be due to projection effects and, to obtain a sample of more
reliable physical structures, we have investigated their red galaxy content, since
it is well know that the presence of a red sequence is conspicuous in large physical groups
and clusters \citep[e.g.][]{Oemler74, Dressler80, Postman84}.

\begin{figure}
\includegraphics[width=7.75cm]{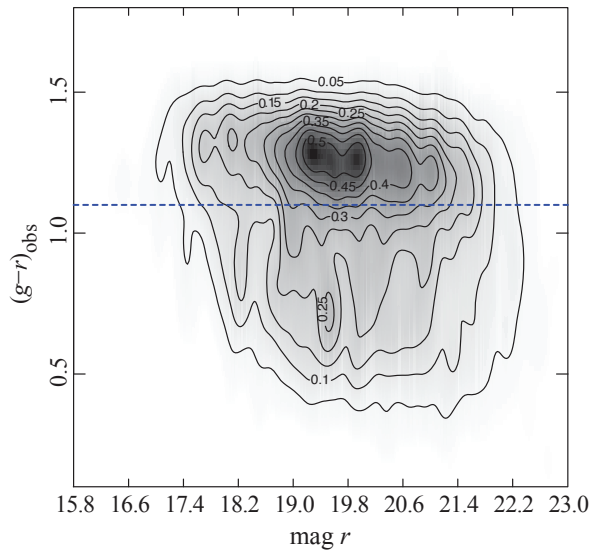}
\caption{$g-r$ colour versus $r$-band magnitude diagram for the galaxies in the 
photometric sample S2. The continuous line at $(g-r) = 1.1$ divides the sample in red and blue galaxies (see text for details).
}
\label{f0}
\end{figure}

We present in Figure \ref{f0} the $(g-r)$ versus $r$ colour-magnitude diagram for galaxies in sample S2. This figure shows a clear concentration of red galaxies.  
In order to make an inventory of red objects, we have divided our sample in two using $(g-r) = 1.1$ as a cutoff  based on the typical colors of galaxies at redshift $z=0.2$ as given by  \citet{Fukugita95}: ellipticals  $(g-r) = 1.31$, lenticulars $(g-r) = 1.13$, Sab galaxies $(g-r)= 1.01$.

The fraction of red galaxies, by this definition, is shown in Table \ref{t1}
for each identified structure.

Based on this information we decided to remove substructures 7, 8 and 9 from our
subsequent analysis, as they all have a fraction of 
red galaxies smaller than $50\%$. We suspect that these substructures may not be \textit{bona fide} 
bound groups and/or clusters of galaxies. We thus kept, for the present study, 6 groups/clusters which 
we considered dense and red enough to be safely considered as substructures of SC0028.
Note that structure 3 appears strong in the spectroscopic redshift map (Figure \ref{f2}), while in the photometric redshift and light 
maps it appears diluted. We have considered this structure significant, since more than $50\%$ of its galaxies are red. 

%----------------------------------------------------------------------------

\begin{figure}
\includegraphics[width=8cm]{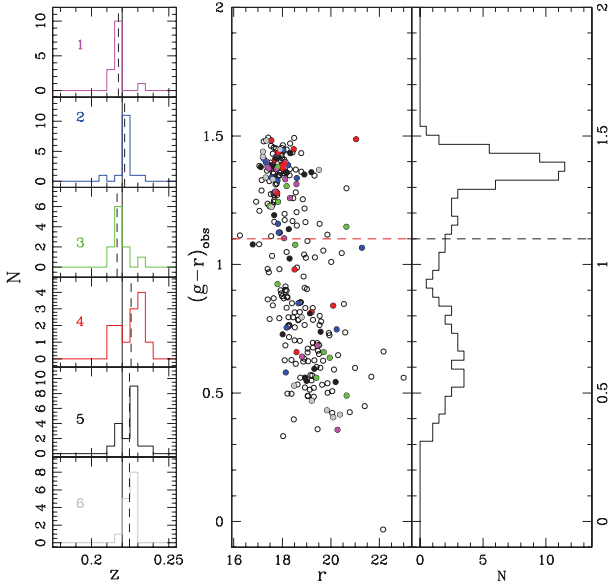}
\caption[]{Results for sample S1. \textsc{Left panel}: spectroscopic redshift distribution of each group 
candidate; dotted lines represent the mean redshift of the group and the continuous line is the mean redshift of the supercluster. 
\textsc{Middle panel}: The $(g-r) \times r$ colour-magnitude diagram. 
Each dot colour represents a structure. Black dots are galaxies in the area not associated to any structure.  
\textsc{Right panel}: The $(g-r)$ distribution. The dashed horizontal line at $(g-r) = 1.1$ on the last two panels indicates our adopted separation between red and blue galaxies.

(A color version of this figure is available in the online journal) }
\label{f4}
\end{figure}

Figure \ref{f4} presents results for the spectroscopic sample, S1.
In the left panel we show the redshift distribution for each substructure. The dotted lines marks its mean redshift 
and the continuous line is the mean redshift of the supercluster. 

On the middle panel we show the $(g-r) \times r$  colour-magnitude diagram of supercluster members in this sample. Dots of a given colour identify members of a specific structure.
On the rightmost panel we show the distribution of the $(g-r)$ colour.

%%%%%%%%%%%%%%%%%%%%%%%%%%%%%%%%%%%%%%%%%%%%%%%%%%%%%%%%%%%%%%%%%%%%%%%%%%%%%%%%%%%%%%%%%
%WEAK LENSING
%%%%%%%%%%%%%%%%%%%%%%%%%%%%%%%%%%%%%%%%%%%%%%%%%%%%%%%%%%%%%%%%%%%%%%%%%%%%%%%%%%%%%%%%%

\section{Weak-Lensing analysis}
\label{sec:lens}

In this section we present a weak-lensing analysis of SC0028, including sample selection and the PSF mapping and correction. We describe the process of 2D mass reconstruction and  fits to the shear profile of individual substructures. 
Notice that the area used for the weak-lensing analysis (based on the CFHT
observation) is significantly smaller than the area used for the spectroscopic
and photometric analysis of the other sections, based on SDSS data 
(see Figure~\ref{f2}).

\subsection{Background sample selection}

The selection of the background galaxies (i.e. those affected by the gravitational lens effect) has been done by identifying 
the regions on a $(g-r) \times (r-i)$ colour-colour diagram where they are abundant and that have the smallest 
contamination by foreground and supercluster galaxies. 
In this section we consider as members of the supercluster the 326 galaxies with spectroscopic redshifts with redshifts between $0.18$ and $0.28$. As a complement, we have also used photometric redshifts in the same interval, from the CFHTLS deep field catalog (hereafter D; Coupon et al. 2009).

\begin{figure*}
\centering
\includegraphics[width=0.49\textwidth,height=0.4\textwidth]{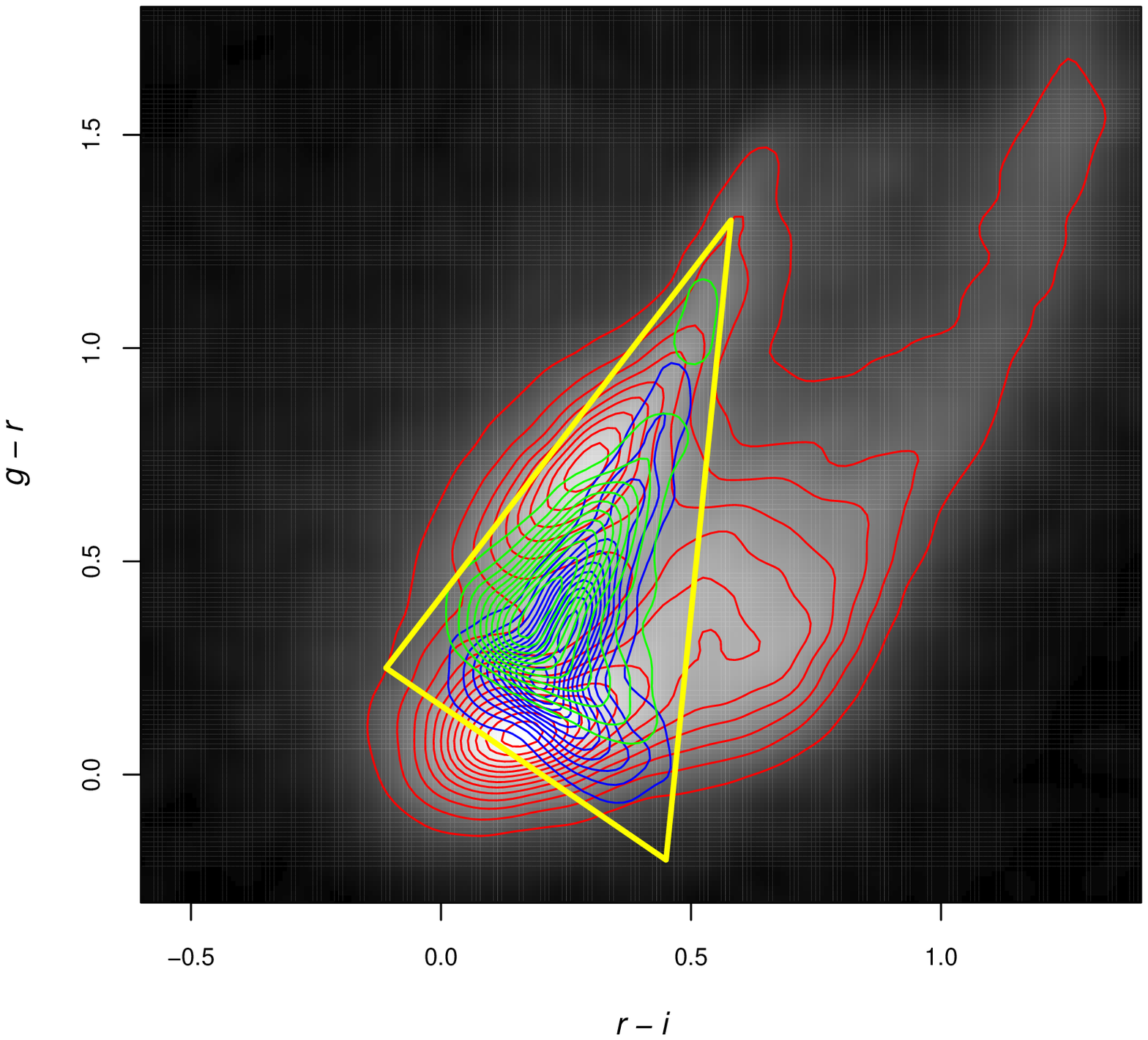}
\includegraphics[width=0.49\textwidth,height=0.4\textwidth]{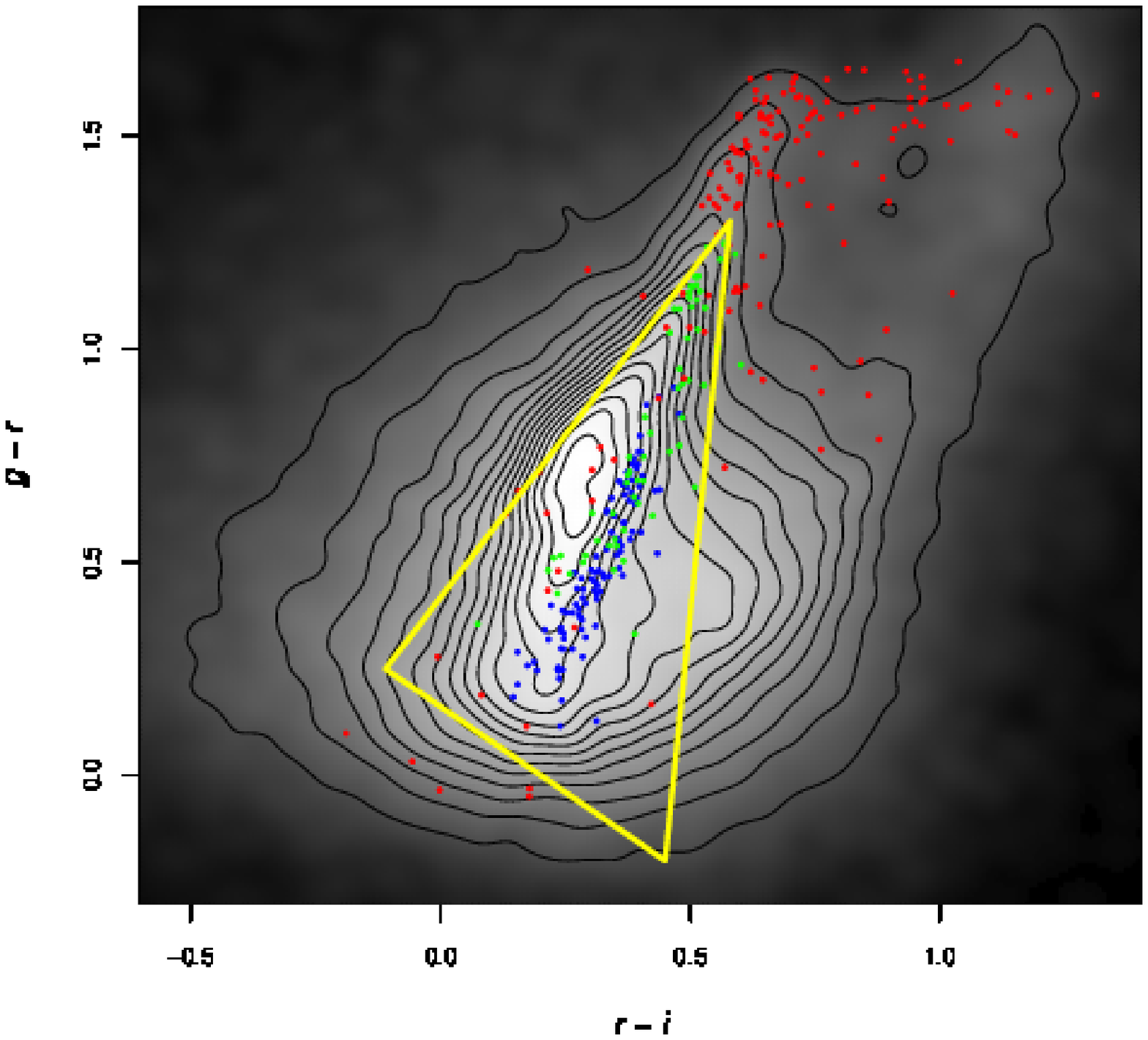}
\caption{
\textsc{Left panel}: Colour-colour diagram identifying the regions that encompasses the galaxies of D catalog (contours) and galaxies. Supercluster galaxies ($0.18\leq z \leq 0.28$) are in green, foreground ones ($z<0.18$) in blue, and background galaxies ($z>0.28$) in red. The yellow triangle encloses the region where most foreground and supercluster galaxies lie.
\textsc{Right panel}: Colour-colour diagram for galaxies in the field of the supercluster SC0028 with spectroscopic redshifts (points).
\label{zsep}}
\end{figure*}

In Figure \ref{zsep} (left panel) we show a $(g-r) \times (r-i)$ colour-colour diagram of the D galaxies (contours) 

with different colours that identify them as foreground ($z<0.18$; blue), supercluster ($0.18\leq z \leq0.28$; green), and background($z>0.28$; red). With
the aid of this information we traced a triangle on this diagram, defined by vertices (-0.11,0.3), (0.58,1.3) and (0.45,-0.16), that selects the regions more
heavily contaminated by  non-background galaxies. All the galaxies outside this selected region are our primary background galaxy sample. Using the D catalogue we have
estimated that the contamination of the background sample by field galaxies is only 0.6\% within the magnitude range $17<r<24.6$. In Figure \ref{zsep} (right panel) we show the colour-colour diagram of the SC0028
field data, with the same triangular region overplotted, where the points represent the spectroscopic redshifts.  Within the same magnitude limits we have, on the CFHT image,
a space density of $\sim20$ arcmin$^{-2}$
candidate background galaxies.

%%%%%%%%%%%%%%%%%%%%%%%%%%%%%%%%%%%%%%%%%%%%%%%%%%%%%%%%%%%%%%%%%%%%%%%%%%%%%%%%%%%%%%%%%%
\subsection{Shape Measurements}

The observed shape of background galaxies is a combination of their intrinsic ellipticity, the shear caused by the lens
plus distortions due to the atmosphere and the telescope+instrument optics (the PSF). The latter can be mapped through the observed shapes of stars (point sources in practice), and we can use it to deconvolve galaxy images.

We measured galaxy and PSF shapes using the \textsc{Im2shape} software \citep{im2shape}, which estimates the ellipticity of astronomical objects by modeling them as a sum of Gaussians with elliptical bases. Stars are modeled as a single Gaussian whereas galaxies require a sum of two of those components, however keeping the centroid ($x_c$, $y_c$) and ellipticity ($\epsilon_{1}$, $\epsilon_{2}$) of both components as the same. The remaining parameters are the amplitudes ($A$) and a parameter related to the area of the base ellipse ($ab$).

\begin{figure}
\centering
\includegraphics[width=7.5cm]{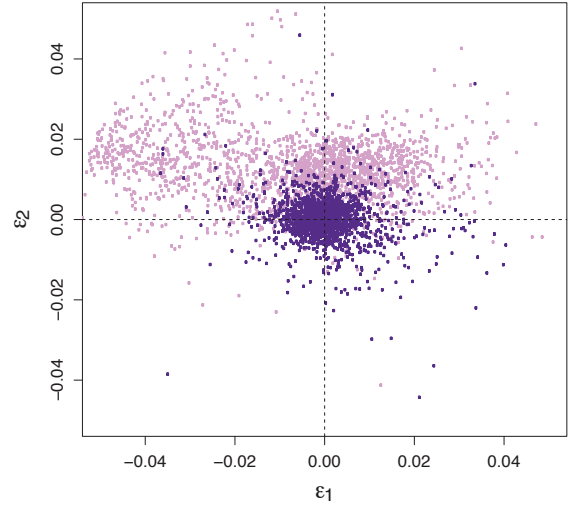}
\caption{Values of components of ellipticity, $\epsilon_{1}$ and $\epsilon_{2}$, before (pink) and after (purple) the correction described in the text, where $\langle\epsilon_{1}\rangle=10.10\times10^{-5}$, $\sigma_{\epsilon_{1}}=7.32\times 10^{-3}$, $\langle\epsilon_{2}\rangle=2.14\times10^{-5}$ and $\sigma_{\epsilon_{2}}=5.21\times 10^{-3} $.}
\label{PSF}
\end{figure}

We selected 1944 stars based on the FWHM ($\sim 0.45$~arcsec) of their PSF in the range ($18<r<22$), where they have good signal-to-noise, are easily separable from galaxies, and show no signs of saturation. From this sample we can estimate the PSF at any point of the image. We did that by smoothing the spatial variation of the relevant PSF parameters ($\epsilon_{1}$, $\epsilon_{2}$ and $ab$) with a Gaussian filter

\begin{equation}
PSF_{par} =\frac{\sum_i w_{i} PSF_{par,i}}{\sum_i w_{i}}\text{,}~~\text{with}~~w_{i}=\exp\left( {\frac{-d_{i}^{2}}{2\sigma^{2}}}\right),
\label{filtro1}
\end{equation}
where $PSF_{par}$ represents one of the relevant parameters at a given point of the image and $d_{i}$ is the projected distance from the point to the star. The free parameter here is $\sigma$. We found a best value of 110~arcsec by minimizing the variance between predicted (interpolated) values with the actual values measured from the stars (See Figure \ref{PSF}).

Next, we applied {\sc Im2shape} to the galaxies, tuning  it to perform a PSF deconvolution using the predicted values for each galaxy position. We then excluded all the galaxies with a composed ellipticity error $\sigma_{\epsilon_{1}}^2 + \sigma_{\epsilon_{2}}^2 > 0.45$, ending up with a sample of 54187 galaxies ($\sim 15$ gal.~arcmin$^{-2}$).

%%%%%%%%%%%%%%%%%%%%%%%%%%%%%%%%%%%%%%%%%%%%%%%%%%%%%%%%%%%%%%%%%%%%%%%%%%%%%%%%%%%%%%%%%%%%%%%%%%%

\subsection{Mass Map}

We reconstruct the 2D density map from the shear information using the {\sc Lensent2} algorithm \citep{lensent1,lensent2}, which creates a convergence ($\kappa$) map through a maximum entropy method from the ellipticity components of each background galaxy $\epsilon_{1}$ and $\epsilon_{2}$ and their respective 
uncertainties. To prevent overfitting, the resultant mass distribution is convolved with a Gaussian kernel. For that we choose an scale of 150~arcsec that yields a significant result, as estimated using the Bayesian evidence, without compromising too much the spatial resolution. 

The convergence is translated in to a physical mass density by multiplying it with the critical lensing surface mass density
\begin{equation}
\Sigma_{crit} = \frac{c^2 }{4 \pi G} \frac{D_s }{D_l D_{ls}}, 
\end{equation}
where $D_l$, $D_s$, and D$_{ls}$ are the angular diameter distances from
the observer to the lens, from the observer to the source, and from the
lens to the source. 

Using the photometric redshift of the deep CFHTLS catalogues we estimated an average ratio $\langle D_s/D_{ls} \rangle$ of 1.39 \citep[See][for a description of the procedure]{cypriano04}. Therefore, we obtained  $\Sigma_{crit} = 3.28\times 10^{15}$ M$_\odot$/Mpc$^{2}$ = 0.685 g/cm$^2$.
The resultant mass distribution is shown in the lower right panel of Figure \ref{f2}.

%%%%%%%%%%%%%%%%%%%%%%%%%%%%%%%%%%%%%%%%%%%%%%%%%%%%%%%%%%%%%%%%%%%%%%%%%%%%%%%%%%%%%%%%%%%%%%%%

\subsection{Individual Masses}
We determined the masses of each substructure in the supercluster by fitting NFW \citep{NFW96,NFW97} model predictions to their shear profiles up to  a radius of $10$ arcmin centered in each optically identified structure. Other values for radius have been tested, however the estimated masses remain unchanged within the statistical errors. The analytical expressions for the radial dependence of the shear of the NFW model can be seen in \citet{Wright00}. 

The NFW profile can be completely defined by two parameters: the mass inside a region with density 200 times the critical density, $M_{200}$, and a concentration parameter $c$. The latter is poorly constrained by our data and thus we fix it at the value of $c=6.79$, which is appropriate for $10^{14} M_\odot$ halos at $z=0.2$ \citep{Prada12}.

We obtain the masses by minimizing the standard statistical misfit between data and model:
\begin{equation}
\chi^{2}=\sum_i{\frac{(\epsilon_{t,i}-g_{t,i})^{2}}{\sigma_{ell}^{2}+\sigma_{m,i}^{2}}} \, ,
\end{equation}
where $\epsilon_{t,i}$ is the measured tangential component of the ellipticity with respect to the substructure centre for the i-th galaxy and $g_{t,i}$ is the model prediction. $\sigma_{ell}$ is an error related to the intrinsic ellipticity of the galaxies, for which we measured a value of $\sim$0.3, and $\sigma_{m,i}$ is associated to the measurement error of ellipticities. The latter is estimated as:
\begin{equation}
\sigma_{m}=\sqrt{\frac{\sigma_{\epsilon_{1}}^{2}+\sigma_{\epsilon_{2}}^{2}}{2}}\, ,
\end{equation}
where $\sigma_{\epsilon_{1}}$ and $\sigma_{\epsilon_{2}}$ are the uncertainties in both components of the ellipticity estimated by {\sc Im2shape} in the fitting process.

In Table~\ref{tab:masses} we show the masses of the four identified substructures which are in the CFHT imaging field. The mass of region 3 is consistent with zero so it does not appear in this table. Region 6 does not appear either because its centre lies outside the CFHT image.

\begin{table}
\centering
\caption[]{Weak-lensing masses of SC0028 substructures with the CFHT imaging field.}
\begin{tabular}{cc} 
\hline
\hline
ID    & $M_{200}$ $(10^{14} M_{\odot})$ \\
\hline
1     &  $0.65 \pm 0.40$ \\ 
2     &  $2.04 \pm 0.58$ \\ 
4     &  $1.21 \pm 0.53$ \\ 
5     &  $1.82 \pm 0.88$ \\
\hline
\hline
\label{tab:masses}
\end{tabular}

\end{table}

%%%%%%%%%%%%%%%%%%%%%%%%%%%%%%%%%%%%%%%%%%%%%%%%%%%%%%%%%%%%%%%%%%%%%%%%%%%%%%%%%%%%%%%%%

\section{Dynamical analysis}\label{sec:dynamical}
In this section we verify if the dynamical behaviour of the substructures identified in the
supercluster are consistent with the collapse scenario described in the Introduction. For
this, we first determine the relative distances and peculiar velocities of these substructures.
We also examine our results with a simple dynamical model.

\subsection{Scaling relations: the Fundamental Plane}\label{sec:scaling}

In this section we determine the Fundamental Plane of early-type galaxies to obtain,
in the next section, estimates of the distances to the substructures identified in the supercluster. 

The Fundamental Plane (FP) is an empirical relation between the central velocity 
dispersion ($\sigma_0$), the physical effective radius ($R_0$) and the mean surface 
brightness ($\mu_e$) within the effective radius \citep{terlevich,dressler,dj}. 
The FP can be represented as:
\begin{equation} \label{fp}
\log_{10}(R_0) = a \log_{10} (\sigma_0) + b \log_{10} (I_e) + c \, ,
\end{equation}

where $I_e$ is the mean effective surface brightness in linear unities, $\mu_e=-2.5 \log_{10}(I_e)$. 

If the FP is to be used as a distance estimator,  it is convenient to adopt a direct fit for the coefficients \citep{Bernardi03}, 
because it minimizes the dispersion in the physical radius $R_0$.   
Another option is by applying orthogonal fits, but this approach is more useful for the study of the global properties of elliptical galaxies or to constrain their underlying physics. Hence, we use here a direct fit for the calibration of the Fundamental Plane to use it as a distance indicator.

Following \citet{Bernardi03} and \citet{hy} we first renormalized the effective 
radii from the SDSS data ($r_{\rm sdss}$) 
using the ratio of of the minor and major galaxy semi-axes ($q_{\rm AB}$) to account for the  ellipticities  of the galaxies in our sample.

Thus, we adopted as our galaxy radius:
\begin{equation}
r = r_{\rm sdss} \sqrt {q_{\rm AB}} ~ .
\end{equation}
This was done to avoid a bias due to the distribution of ellipticities \citep{Bernardi03}.

Since the SDSS uses a fixed fiber size, the fibers cover different galaxy physical areas at 
different distances.  
Therefore, we need to take this 
into consideration and correct the velocity dispersion for the spectroscopic galaxies 
in our sample.

The aperture corrections for early type galaxies were calculated by \citet{Jorgensen95} and 
\citet{Wegner99} as follows:
\begin{equation}
\sigma_0= \sigma_{\rm sdss}(\frac{r_{\rm fiber}}{r})^{0.04}\, ,
\end{equation}
where $r$ is the corrected radius (see above) and $r_{fiber}$ is 1 arcsec  \citep{Ahn12}.

Finally, the surface brightness in a circle of radius $r$ is defined as:
\begin{equation}
\mu_0= m + 2.5 \log_{10}(2 \pi r^2) \, ,
\end{equation}
where $m$ is the apparent magnitude corrected by extinction and $k$-correction. In this work, $k$-corrections were calculated using the publicly available software \textsc{k-correct} v4.2 of \citet{Blanton07}.  The effective radius $R_0$ is in physical units of $h_{70}^{-1}$~kpc.

%========================================================
\begin{figure}
\includegraphics[width=8.6cm]{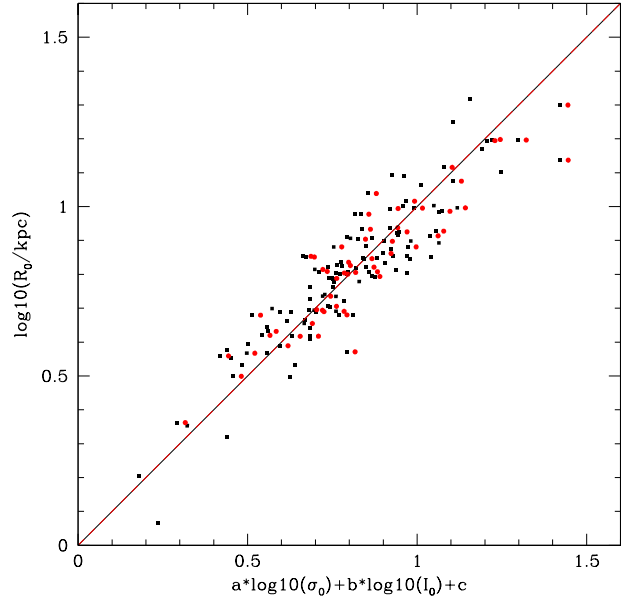}
\caption[]{ 
The Fundamental Plane relation. Black box symbols correspond to all red galaxies ($(g-r)>1.1$) within $\Delta z < 0.02$ of
the supercluster mean, whereas red symbols show galaxies belonging to the supercluster substructures. The dotted and dashed lines are the best linear fittings to the black and red symbols, respectively. The fittings are almost identical, making the distinction of the two lines difficult.
The rms dispersions are 0.094 and 0.079 for the black and red dots, respectively.}
\label{f8}
\end{figure}

In this subsection and in the next, we work with the S1 sample. Since we are interested here in early-type galaxies, 
we consider only red galaxies, those with $(g-r) > 1.1$.  

We verified that the resulting sample is consistent with the morphological classification of GalaxyZoo (\citet{glzoo}), available for 52 out of the 57 galaxies of our sample. We used for this end the assigned probabilities of being ellipticals ($P(E)$). For the 52 
galaxies we found that 49 have $P(E) > 0.8$, 2 $P(E) > 0.71$ and 1 has $P(E) \sim 0.2$.  
Additionally, the distribution of apparent axial-ratios is consistent with that of {\it bona fide} elliptical galaxies extracted from the Third Reference Catalogue of Bright Galaxies (RC3) by G. de Vaucouleurs, A. de Vacouleurs, H.G. Corwin, R.J. Buta, P. Fouque, and G. Paturel (\citet{rc3}). We found  $\sim$ 10\% of our galaxies having $(b/a) < $ 0.6, where $a$  and $b$ are the apparent major and minor axis, similar to the RC3 sample. We conclude that our sample of galaxies should be less than 10\% contaminated by non-elliptical galaxies, with no impact on the FP fitting.

Figure \ref{f8} shows the best FP fit. The black box dots represent the 123 red galaxies of the spectroscopic sample 
with reliable central velocity dispersion 
measurements and within $\Delta z < 0.02$ of the mean supercluster redshift. The red dots represent 57 galaxies within the six substructures 
we identified.  
The best fit parameters for the whole supercluster 
and for galaxies in the substructures are virtually indistinguishable, with
$a=1.035 \pm 0.036$, and $b=-0.775 \pm  0.031$ and a rms scatter of  0.0941 and 0.0787 for the whole and substructure samples, 
respectively. These values  agree well with those obtained by \citet{Saulder13} for the SDSS-DR8.

\subsection{Determination of distances and peculiar velocities}\label{sec:dists_velocities}

In this Section, we first estimate relative distances to each substructure and, together with mean
radial velocities, we estimate their peculiar radial velocities. Since we have on average only 10 galaxies in the S1 sample per substructure (see Table \ref{t2}), we do not use the FP to directly measure distances to the supercluster components; instead, the zero-point offset of each sustructure with respect to the overall fit can be used to find relative distances between them, keeping the coefficients $a$ and $b$ of Eq. \ref{fp} fixed. The values of the fitting coefficients are shown in Table \ref{t2}, for the supercluster as a whole and for the substructures.

Following \citet{Pearson14} and \citet{Batiste13}, we assume that the 
structures are, on average, at rest with respect to the CMB. Specifically, we do not assume any peculiar motion for the supercluster centroid.  Individual substructures offsets are shown in Table \ref{t2}.

%===============================================================
\begin{table}

\caption{
Results for the best fit of the the Fundamental Plane for the supercluster as a whole
and the substructures. Column (1) shows the identification of the structure in consideration. Column (2) presents
the Fundamental Plane coefficients (see text for more details), with errors calculated via a boostrapping procedure. Column (3) is the $rms$ dispersion in the fits for the supercluster as a whole and the substructures. Column (4) gives the number of galaxies used for the FP fitting.}

\begin{center}\begin{tabular}{cccc}
  \hline
  \hline
 Sample id &   & $rms$ & N \\
\hline
SC   & a=1.035  $\pm$ 0.036   & 0.0787 & 57 \\ 
        & b=-0.775 $\pm$ 0.031   &      \\ 
        & c=-9.105 $\pm$ 0.006   &     \\ 
\hline
1    & c=-9.108 $\pm$ 0.015  & 0.055 &11\\
\hline
2    & c=-9.103 $\pm$ 0.012  & 0.061 &10\\ 
\hline
3    & c=-9.098 $\pm$ 0.014  & 0.062 &8\\ 
\hline
4    & c=-9.127 $\pm$ 0.014  & 0.064 &9\\ 
\hline
5    & c=-9.127 $\pm$ 0.012  & 0.056 &10 \\ 
\hline
6    & c=-9.101 $\pm$ 0.017  & 0.065 &9\\ 
\hline
\label{t2}
\end{tabular}
\end{center}
\end{table}

The error of individual galaxy distances estimated from the FP is $\Delta = \ln(10) \times rms$; since the dispersion of the FP relation is $\sim 0.079$ (Table \ref{t2}), $\Delta$ is around 18\%, in agreement with the values obtained by \citet{Pearson14} and \citet{Batiste13}. For individual distances, the error percentage is reduced, assuming a Poissonian distribution, by  $\Delta/\sqrt{N}$ (where $N$ is the number of galaxies in the substructure) and is tipically 6\%.

The comoving distances were calculated using the difference of zero points and converted to 
redshift by the approximation \citep{Peebles93}:
\begin{equation}
D=\frac{cz}{H_0}(1-z\frac{1+q_0}{2}) \approx 4283 (1 -0.225 z) \, z  h_{70}^{-1}~\mbox{Mpc}\, ,
\end{equation}
which is appropriate at $z \sim 0.22$. We assume $q_0 = -0.55$. The peculiar velocities, $v_p$, were measured for each 
substructure using the difference between the redshift obtained  with the Fundamental Plane ($z_{FP}$) and the mean
spectroscopic redshift of the supercluster ($z_m$):
\begin{equation}
v_p=c\left ( \frac{z_m-z_{FP}}{1+z_{FP}} \ \right)
\end{equation} 
A negative velocity indicates that the substructure moves towards us, while a positive velocity 
indicates that it moves away from us.

A summary of the main results of this section is given in Table \ref{t3}.

\begin{table}
\caption[]{
Fundamental Plane mean redshifts and peculiar velocities for each substructure.
Column (1) gives the identification of the substructures. Column (2) gives the cluster redshift determined from the
Fundamental Plane and Column (3) gives the derived peculiar velocity.
Column (4) gives the error in the distances determined from the $rms$ dispersion of the global fit and
the number of objects in each substructure. 
}
\begin{center}\begin{tabular}{cccc}
\hline
\hline
 Id & $z_{FP}$ & $V_p$[km/s] & err[\%]  \\
\hline
1 & 0.219 &  -489.810 & 4  \\
2 & 0.217 &  1025.577 & 4  \\
3 & 0.215 &  269.444  & 5  \\
4 & 0.228 &  -611.016 & 5  \\
5 & 0.228 &  -815.231 & 4  \\
6 & 0.216 &  2036.244 & 4  \\
\hline
\hline
\label{t3}
\end{tabular}
\end{center}
\end{table}

\subsection{SC0028 as a collapsing structure}

Now we test whether or not the substructures we identified are in a process of collapse towards the supercluster barycentre. The most identifiable feature of the process is a simple redshift space distortion. Substructures which are physically closer to the observer than the supercluster centre would feel its gravitational pull towards it, and thus will appear as having positive peculiar line-of-sight velocities (i.e., $V_p$ points away from the observer, adding to the receding velocity.). The opposite is expected in case of substructures which are physically farther than the supercluster barycentre.

To test this hypothesis, we have used the FP distances and peculiar velocities determined in the previous subsection. The average distance of SC0028 is $\sim 1090.50 ~ h_{70}^{-1}$ Mpc so, disregarding statistical uncertainties, substructures 2, 3 and 6 would be in front of the supercluster and thus should have positive peculiar velocities, while substructures 1, 4 and 5 would be behind and are supposed to have negative  peculiar velocities.

% ==================================================================================
\begin{figure*}
\includegraphics[width=7.2cm]{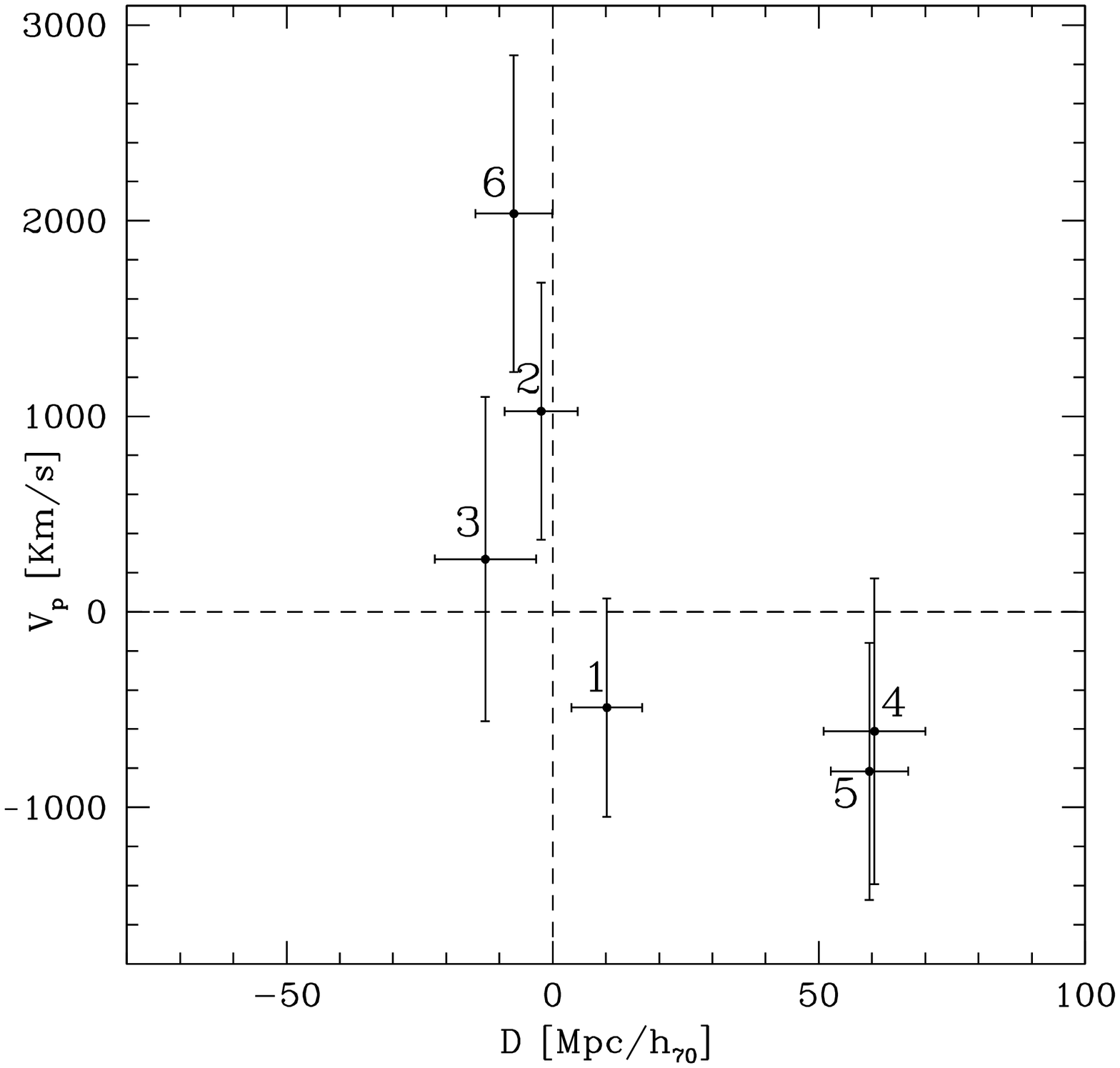}\kern1em
\includegraphics[width=9cm]{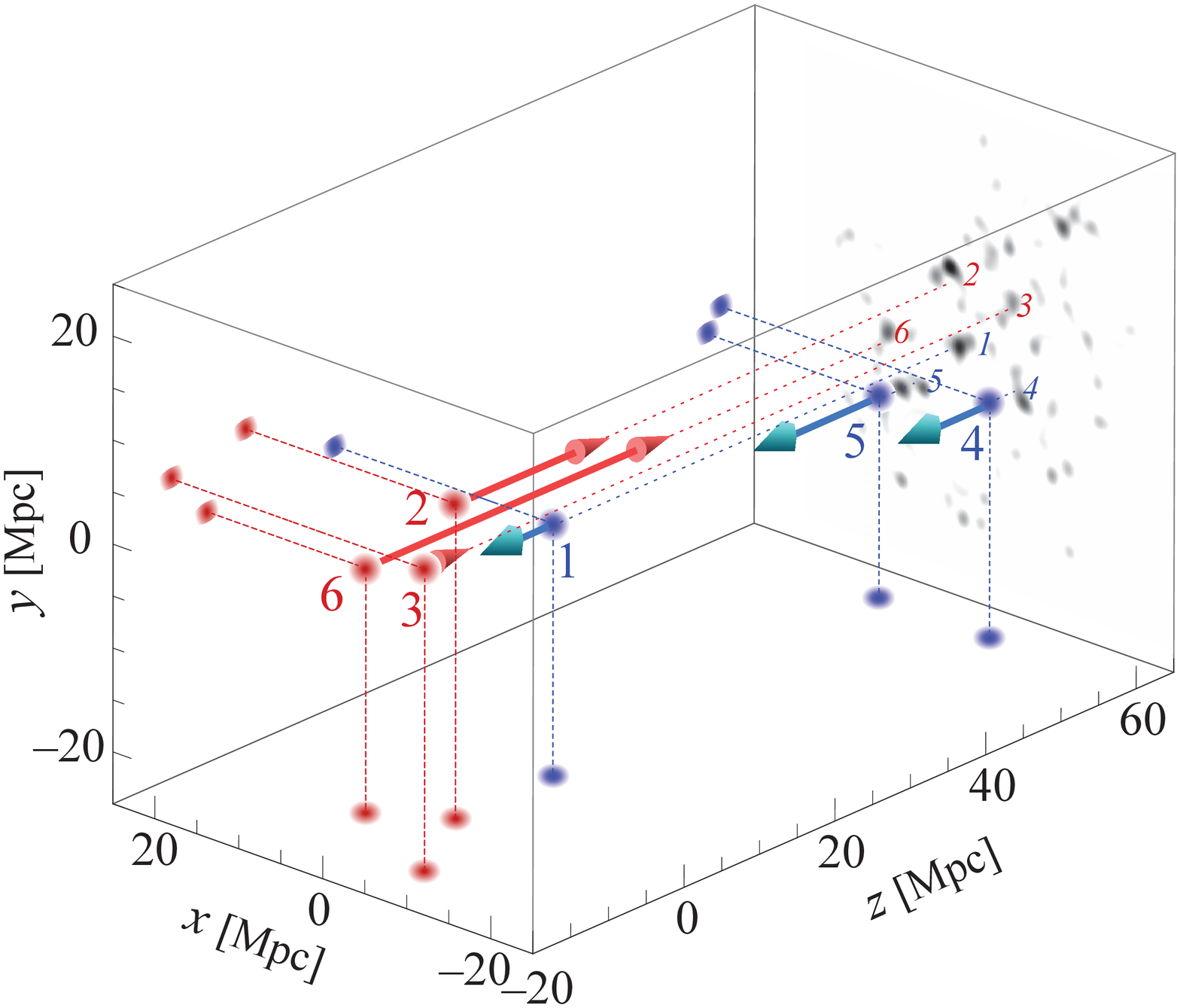}
\caption[]{\textsf{Left panel}: Distances relative to the supercluster center \textit{versus} peculiar velocities for the six substructures,
using the Fundamental Plane method. \textsf{Right panel}: The 3-dimensional spacial distribution of SC0028 substructures also showing the LOS velocity vector. The length of the $V_p$ vectors are in arbitrary units, proportional to the velocities derived using the Fundamental Plane.
}
\label{f12}
\end{figure*}
% ==================================================================================

Assuming a Normal distribution of probabilities for the peculiar velocities and distances, it is easy to estimate the total probability that a particular cluster/group is in one of the two expected quadrants in the $D_r \times v_p$ space, {\it i.e.} the probability of having positive peculiar velocities and negative relative distances, plus the probability of having negative peculiar velocities and positive relative distances.
Assuming independence between those quantities and their errors,  we find the following probabilities for each substructure be indeed in the expected quadrant where it is found: from substructure 1 to 6 we obtain, respectively, 59.4, 53.5, 53.2, 65.2, 73.2, 65.3\%.
Combining all these p-values using the Fisher method \citep{H&O85}, we obtain a probability of 92.2\%. 
This number can be interpreted as a moderate evidence for the detection of the infall of these substructures.

Clearly most of the signal comes from substructures 4 and 5, which, in particular, presents no ambiguity regarding being on the far side of the supercluster. The overall  probability is fairly insensitive to the exclusion of the less certain substructure 3.

The left panel of Fig.~\ref{f12} shows the phase space $D$--$V_p$ for the 6 substructures of SC0028. The corresponding three dimensional structure is depicted on the right panel of Fig.~\ref{f12}, where the present collapsing state of the supercluster can be appreciated. 
The errors in $V_p$ were calculated through error propagation considering the error in the distance of each substructure.

\subsection{Spherical collapse model}

In order to estimate the supercluster mass, we employ the spherical collapse model in an expanding universe, that describes the free-fall evolution of a shell enclosing some mass $M$ \citep{Gunn72,Lahav91}. Albeit simple, the spherical collapse model has been extensively studied and used to estimate the mass of superclusters by several authors \citep[e.g.][]{Dunner06,Araya-Melo09,Batiste13,Pearson14}.

Here, we follow the time evolution of a spherical shell by solving simultaneously the motion equation \citep{Peebles03},
\begin{equation}
\frac{d^2 r}{d t^2} + \frac{G M}{r} - H^2(z) \Omega_\Lambda r = 0 \, ,
\end{equation}
and the background Friedman equation,
\begin{equation}
\frac{d z}{d t} + (1 + z) H(z) = 0 \, ,
\end{equation}
with 
$$ H(z) = H_0 \sqrt{\Omega_\Lambda + \Omega_M (1+z)} \, ,$$

\begin{figure}
\includegraphics[width=8.5cm]{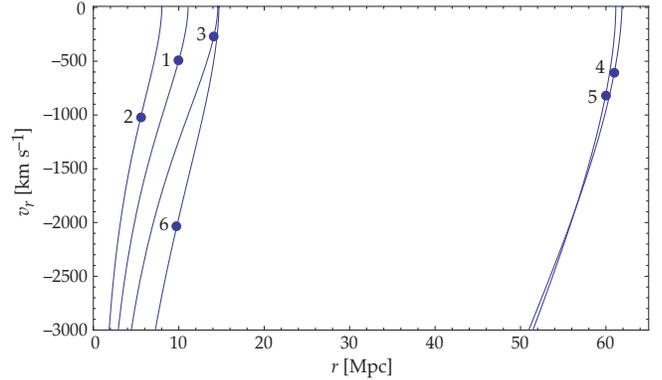}
\caption[]{Free fall trajectories for the six substructures in SC0028 in phase-space. The inicial conditions are marked as large dots, with the corresponding numbers.}
\label{fig:freefall}
\end{figure}

For each one of the six substructures, we have solved the above differential equations using, for the inicial conditions, $z(0) = 0.22$, and $r(0)$ and $v_{r}(0)$ corresponding to each group as in Fig.~\ref{f12}. Notice that $v_{r}(0)$ is always negative since the velocity vectors for all substructures point towards the centre of SC0028.

After setting the inicial conditions and cosmological parameters (the standard $\Lambda$CDM model), there still one unknown in the equations since we do not know \textit{a priori} the mass of the supercluster. We have searched interactively for a solution with the present observed values for $r(0)$ and $v_{r}(0)$ and with the smallest density contrast, $\delta_{\rm sc} \equiv \overline{\rho}_{\rm sc}/\rho_c$, where $\rho_c$ is the critical density at the supercluster redshift $z=0.22$. This yields a lower limit estimate for the supercluster mass.

We have integrated the motion equations forward and backward in time, so that we had the full trajectories from the turn-around point onward. Figure~\ref{fig:freefall} shows these trajectories in phase-space and Table~\ref{tab:freefall} summarizes the numerical integration results.

\begin{table}
\centering
\caption[]{Initial conditions, $r(0)$ and $v_{r}(0)$, and results of the integration of the spherical collapse equation. $\delta_{\rm sc}$ is the mean density contrast of the supercluster with respect to the critical density. $t_{\rm TA}$ and $r_{\rm TA}$ are the time of turn-around and the radial distance at that time.}
\label{tab:freefall}
\begin{tabular}{lrrrrrr}
\hline
\hline
Obj.  &  $r(0)$  & $v_{r}(0)$ & $\delta_{\rm sc}$ & $M[<r(0)]$ & $t_{\rm TA}$ & $r_{\rm TA}$ \\
 Id   &  [Mpc]   &   [km/s]     &                   & $[10^{15} M_\odot]$ &  [Gyr]  &  [Mpc] \\
\hline
1 & 10.0 &  -490 &   6.1  &   4.3  &  -5.7  &  11.1 \\
2 & 5.5  & -1026 &  21.6  &   2.6  &  -6.5  &  8.0 \\
3 & 14.1 &  -269 &   3.6  &   7.2  &  -5.5  &  14.6 \\
4 & 61.0 &  -611 &   2.8  &   445  &  -3.6  &  61.9 \\
5 & 60.0 &  -815 &   3.2  &   489  &  -5.7  &  61.0 \\
6 &  9.8 & -2036 &  24.9  &  16.4  &  -7.4  &  14.7 \\
\hline
\hline
\end{tabular}
\end{table}

Figure~\ref{fig:freefall} shows that substructures 1, 2, 3 and 6 have nearby trajectories in phase space, while substructures 4 and 5 have very similar trajectories, but distinct from the other four. The latter pair seems to have about the same turn-around radius, compatible with an enclosing total mass of $\sim 4.5$--$4.9 \times 10^{17} M_\odot$. This large mass, however, corresponds to a mean density contrast of only $\sim 3$ above the critical background density at $z=0.22$. 

The case of substructures 1, 2, 3 and 6 is more complex. Objects 2 and 6 suggest a high density contrast, $\sim 22$--25, while objects 1 and 3 suggest a low density contrast, in agreement with substructures 4 and 5 (between $\delta_{\rm sc} \sim 4$ and 6). On the other hand, substructures 3 and 6 have almost the same turn-around radius, $r_{\rm TA} \sim 14\,$Mpc.

Given the uncertainties in the distance determination and our lack of information on the transverse component of the substructures peculiar velocities -- we have used the line-of-sight peculiar velocity as a rough estimate of the radial peculiar velocity, $v_{r}$ -- we can only have an order of magnitude estimate on the supercluster mass. At a radius of 10~Mpc, the free-fall motion suggests that mass should lie between $\sim 4$ and $16 \times 10^{15} M_\odot$. 

The weak lensing masses of the individual substructures are of the order of 1--$2 \times 10^{14} M_\odot$ (see Table.~\ref{tab:masses}). Within 10~Mpc from the supercluster center, there are 3 to 4 of such substructures that adds up to about $\sim 3 \times 10^{14}$ (without substructure 6, which lies outside the footprint of CFHT). Therefore, most of this supercluster mass budget should be in the form of lower than $\sim 5 \times 10^{13} M_\odot$ galaxy groups. That is below our detectable level with the present data.

%%%%%%%%%%%%%%%%%%%%%%%%%%%%%%%%%%%%%%%%%%%%%%%%%%%%%%%%%%%%%%%%%%%%%%%%%%%%%%%%%%%

\section{Summary and conclusions}\label{sec:conclusion}

In this work we have analyzed the gravitational dynamics of the supercluster SC0028-005 
with spectroscopic and photometric data from SDSS-DR10. This supercluster was selected from
the supercluster catalogue of \citet{Basilakos03}, where it was classified as a filamentary structure. 

We have chosen an area of 1.2 square degrees around the supercluster and proceeded to 
identify substructures using density maps and red-sequence analysis.
We identified 9 substructures throughout the area, but 3 of them were discarded because 
their fraction of red galaxies was less than $50\%$.

To test the collapse scenario for supercluster evolution, we have determined the distances and peculiar velocities for each
substructure in the supercluster. 
In order to study the dynamics of these substructures, the Fundamental Plane was used as a distance indicator.
For the spectroscopic sample, each substructure has, on average, 10 galaxies. This number 
is too small to determine the local Fundamental Plane parameters. Instead we have used the globally fitted slopes and then estimated zero point 
offsets for each of the six substructures.

In order to obtain the individual distances and peculiar velocities, we 
considered that the supercluster's centroid does not have peculiar movements
and, knowing the supercluster's distance beforehand, we computed the individual 
distances for each structure.

The resulting picture for SC0028 is of three substructures approaching the supercluster barycentre from the far side and three from the near side, based
on the computed empirical relations with photo-spectroscopic data. 
The spatial distribution and peculiar velocities of the detected members of SC0028 support the collapsing supercluster scenario. 
Assuming that the dynamics may be described by a simple free-fall spherical collapse model, the overall density contrast is $\delta \sim 3$ 
inside a radius of 60~Mpc. In the inner region, inside 10~Mpc, the mass should be $4 - 16 \times 10^{15} M_\odot$. Most of this should be in the form of low mass ($M \la 5 \times 10^{13} M_\odot$) substructures.

%%%%%%%%%%%%%%%%%%%%%%%%%%%%%%%%%%%%%%%%%%%%%%%%%%%%%%%%%%%%%%%%%%%%%%%%%%%%%%%%%%%%%%%%%%%%%%

\section{Acknowledgments}

We acknowledge financial support form CAPES/Cofecub French-Brazilian cooperation, CNPq and FAPESP.

Funding for SDSS-III has been provided by the Alfred P. Sloan Foundation, the 
Participating Institutions, the National Science Foundation, and the U.S. Department 
of Energy Office of Science. The SDSS-III web site is http://www.sdss3.org/.

SDSS-III is managed by the Astrophysical Research Consortium for the Participating 
Institutions of the SDSS-III Collaboration including the University of Arizona, the
 Brazilian Participation Group, Brookhaven National Laboratory, Carnegie Mellon University, 
University of Florida, the French Participation Group, the German Participation Group, 
Harvard University, the Instituto de Astrofisica de Canarias, the Michigan State/Notre 
Dame/JINA Participation Group, Johns Hopkins University, Lawrence Berkeley National 
Laboratory, Max Planck Institute for Astrophysics, Max Planck Institute for Extraterrestrial 
Physics, New Mexico State University, New York University, Ohio State University, 
Pennsylvania State University, University of Portsmouth, Princeton University, 
the Spanish Participation Group, University of Tokyo, University of Utah, Vanderbilt 
University, University of Virginia, University of Washington, and Yale University.

This work is based (in part) on data products produced at the TERAPIX data center located at the Institut d'Astrophysique de Paris and on observations made with ESO Telescopes at the La Silla Paranal
Observatory (Chile) under programs ID 077.A-0262 and 085.A-0116.

%%%%%%%%%%%%%%%%%%%%%%%%%%%%%%%%%%%%%%%%%%%%%%%%%%%%%%%%%%%%%%%%%

\label{lastpage}

\end{document}